\documentclass[
prc,%
10pt,%
final,%
notitlepage,%
oneside,%
twocolumn,%
nobibnotes,%
nofootinbib,
superscriptaddress,%
floatfix,%
floatfix,%
showkeys,%
showpacs]%
{revtex4}
\usepackage{color}
\usepackage{amsfonts}
\usepackage{amsbsy}
\usepackage{mathrsfs}
\usepackage{graphicx}
\def\lsim{\mathrel{\rlap{
\lower4pt\hbox{\hskip-3pt$\sim$}}
    \raise1pt\hbox{$<$}}}     
\def\gsim{\mathrel{\rlap{
\lower4pt\hbox{\hskip-3pt$\sim$}}
    \raise1pt\hbox{$>$}}}     
\def\scr#1{\mbox{\scriptsize #1}}
\begin{document}
\title{	
Vorticity and Particle Polarization in Relativistic Heavy-Ion Collisions
} 
\author{Yu. B. Ivanov}\thanks{e-mail: yivanov@theor.jinr.ru}
\affiliation{Bogoliubov Laboratory for Theoretical Physics, 
Joint Institute for Nuclear Research, Dubna 141980, Russia}
\affiliation{National Research Nuclear University "MEPhI", 
Moscow 115409, Russia}
\affiliation{National Research Centre "Kurchatov Institute",  Moscow 123182, Russia} 
\author{V. D. Toneev}
\affiliation{Bogoliubov Laboratory for Theoretical Physics, 
Joint Institute for Nuclear Research, Dubna 141980, Russia}
\author{A. A. Soldatov}
\affiliation{National Research Nuclear University "MEPhI",
Moscow 115409, Russia}
\begin{abstract}
We review studies of vortical motion 
and the resulting global polarization of $\Lambda$ and $\bar{\Lambda}$ hyperons
in heavy-ion collisions, in particular, within 3FD model. 
3FD predictions for  
the global midrapidity polarization in the FAIR-NICA energy range are presented. 
The 3FD simulations indicate that energy dependence of the observed global polarization of 
hyperons in the midrapidity region is a consequence of the decrease 
of the vorticity in the central region with the collision 
energy rise because of pushing out the vorticity field into  
the fragmentation regions. 
At high collision energies this pushing-out 
results in a peculiar vortical structure consisting of two vortex rings: 
one ring in the target fragmentation region and another one in 
the projectile fragmentation region with  
matter rotation being opposite in these two rings. 
\end{abstract}
\maketitle

\section{Introduction}

Strongly interacting matter characterized by extremely
high baryon and energy densities is created in heavy ion collisions 
at relativistic energies. This matter demonstrates strong collective behavior 
that is well described by relativistic hydrodynamics. The collective behavior
is manifested in directed and elliptic flow \cite{Voloshin:1994mz,Voloshin:2008dg}
that is comprehensively studied 
both experimentally and theoretically for a long time.

\begin{figure}[htb]
\includegraphics[width=8.2cm]{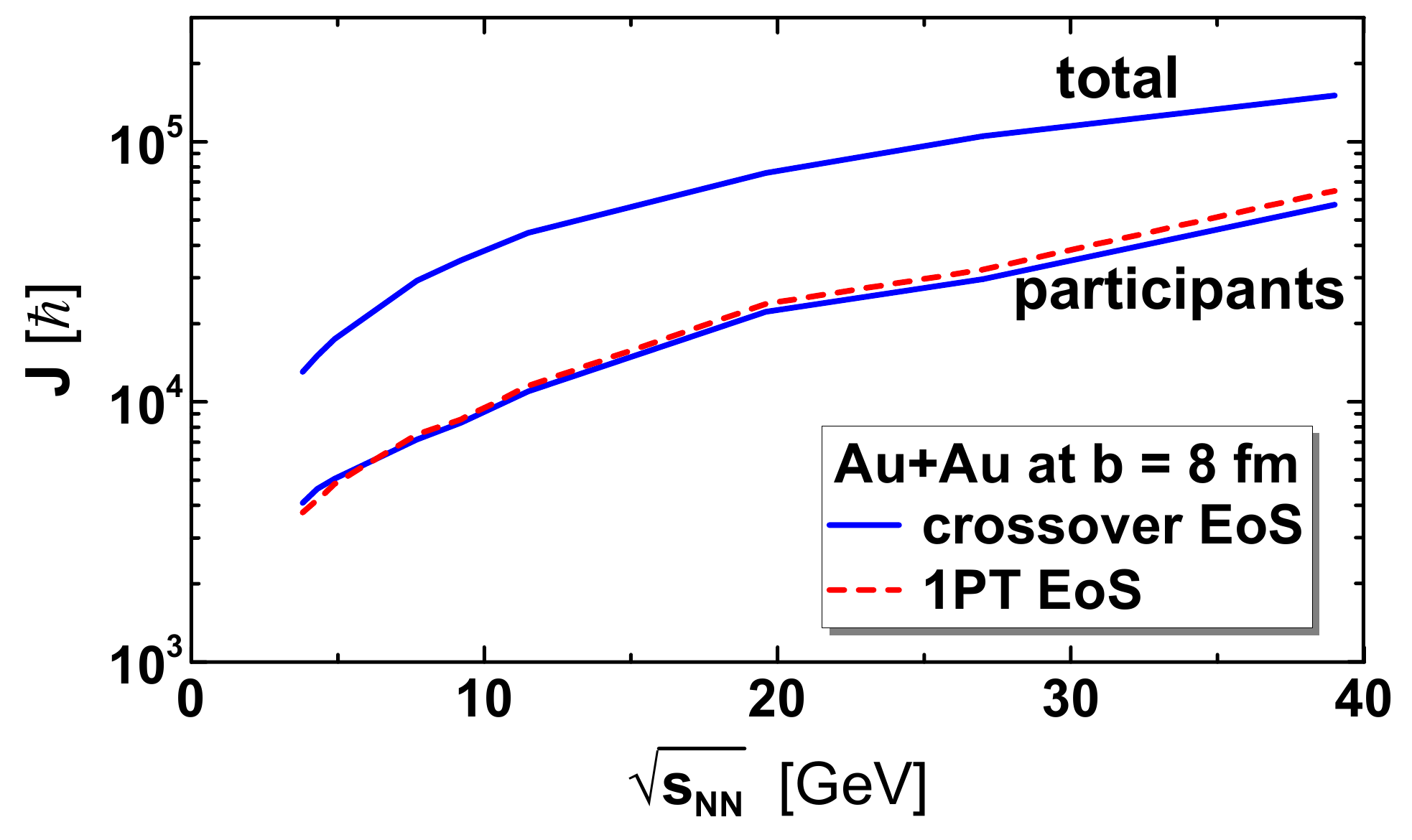}
 \caption{
The total angular momentum (conserved quantity) and  
the angular momentum accumulated in the participant 
region 
in semi-central ($b=$ 8 fm) Au+Au collision as functions of $\sqrt{s_{NN}}$. 
Calculations are done within the 3FD model with the 1PT and crossover EoS's. 
}
\label{fig1}
\end{figure}

Non-central heavy-ion collisions at high energies are also characterized by a huge 
global angular momentum. This is illustrated in Fig. \ref{fig1}, where the total 
angular momentum ($J_{\rm{total}}$) achieved in Au+Au collisions at impact parameter $b=$ 8 fm is 
plotted as a function of the center-of-mass collision energy $\sqrt{s_{NN}}$.
The calculations were performed within the model of the three-fluid dynamics (3FD) \cite{3FD} 
with two versions of the equation of state (EoS) involving 
deconfinement transition \cite{Toneev06}, 
i.e. a first-order-phase-transition (1PT) EoS and a crossover EoS. 
As seen, $J_{\rm{total}}$ rapidly rises with the collision energy, exceeding   
the value of $10^5 \hbar$ at $\sqrt{s_{NN}}>$ 25 GeV. 
It is independent of the used EoS. 
However, only a part of the total angular momentum is accumulated in the participant 
region, i.e. in the overlap region of the interacting nuclei, 
which is of prime interest for us.  
As seen from Fig. \ref{fig1}, 25--30\% of the total angular momentum is deposited into 
the participant matter in the Au+Au collisions at $b=$ 8 fm, which is also a huge amount.  
The participant angular momentum depends, though weakly, on the EoS 
because the overlap region of the interacting nuclei
expands in the course of the collision, including 
more and more former spectators.  
The dynamics of this expansion depends on the EoS.

As the angular momentum is accumulated in the participant region, the motion 
of the matter becomes vortical. In non-relativistic case it is characterized 
by the non-relativistic vorticity  
   \begin{eqnarray}
   \label{nr-vorticity}
\mbox{\boldmath$\omega$} = \frac{1}{2}\mbox{\boldmath$\nabla$} \times {\bf v},  
   \end{eqnarray}
where ${\bf v}$ is a conventional collective 3-velocity of the matter. 
The relativistic generalization of this vorticity is 
   \begin{eqnarray}
   \label{rel.vort.}
   \omega_{\mu\nu} = \frac{1}{2}
   (\partial_{\nu} u_{\mu} - \partial_{\mu} u_{\nu}), 
   \end{eqnarray}
where $u_{\mu}$ is a collective local four-velocity of the matter.

The vortical motion significantly affects the evolution of the system. 
However, theoretical and experimental studies of the relevant effects 
began relatively recently. In particular, in Ref. \cite{Liang:2004ph}
it was suggested that parton interaction in non-central
heavy-ion collisions leads to a global quark polarization
along the direction of the global angular momentum. 
This global polarization is essentially a local manifestation of
the global angular momentum of the colliding system
through spin-orbital coupling \cite{Betz:2007kg,Gao:2007bc}.

This phenomenon of the global polarization along the total angular momentum
is closely related to the Barnett effect \cite{Barnett:1915} 
(magnetization by rotation). The Barnett effect  consists in 
transformation of a fraction of the orbital angular momentum associated 
with the body rotation  into spin angular momenta
of the atoms, which, on the average, are directed 
along the orbital angular momentum. Because of the proportionality
between the spin and magnetic moment, this polarization results in 
a magnetization of the rotating body.

\begin{figure}[htb]
\includegraphics[width=8.2cm]{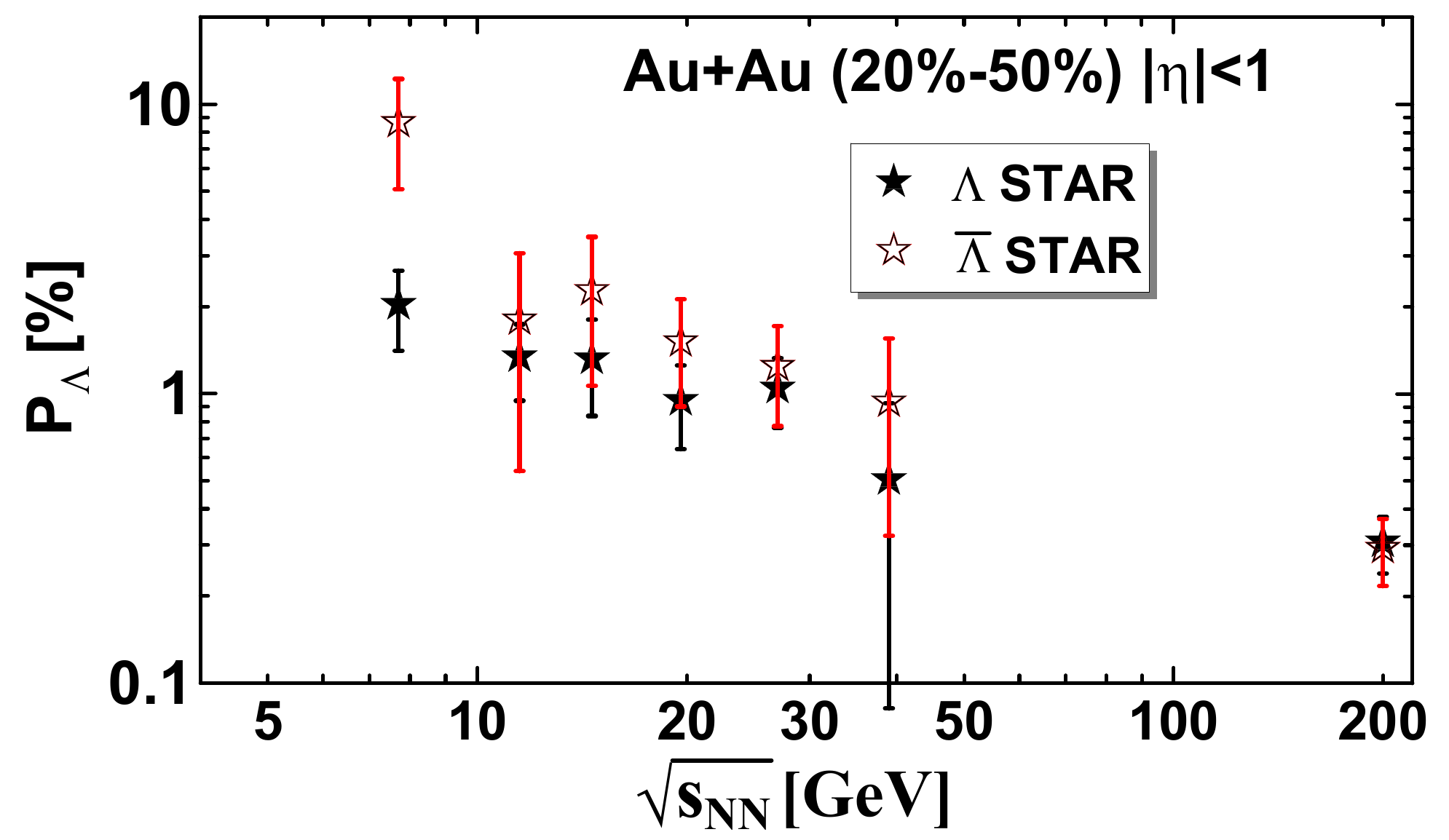}
 \caption{
 Global polarization of $\Lambda$ and $\bar{\Lambda}$ as a function of 
the collision energy $\sqrt{s_{NN}}$ for 20-50\% centrality Au+Au collisions. 
}
\label{fig1.2}
\end{figure}

Measurement of the polarization of hadrons produced in heavy-ion collisions 
is not an easy task by itself. The most straightforward way to detect the global 
polarization in relativistic nuclear collisions is based on measuring polarization
of $\Lambda$ hyperons because they are so-called self-analyzing particles. 
The $\Lambda$ hyperons
decay weakly violating parity: 
$\Lambda \longrightarrow p + \pi^-$ and $\bar{\Lambda} \longrightarrow \bar{p} + \pi^+$. 
In the $\Lambda$ rest frame the daughter proton is predominantly
emitted along the $\Lambda$ polarization (${\bf P}_\Lambda^*$):
   \begin{eqnarray}
\label{L-decay}
   \frac{dN}{d \cos\theta^*} = \frac{1}{2} (1+ \alpha_\Lambda {\bf P}_\Lambda^* \cos\theta^*) 
   \end{eqnarray}
where $*$ means $\Lambda$'s rest frame, $\alpha_\Lambda=-\alpha_{\bar{\Lambda}}=$ 0.642 is 
the $\Lambda$ decay constant \cite{Tanabashi:2018oca}, and $\theta^*$ is the angle 
of the emitted proton. However, first measurements of the 
global polarization of  and $\bar{\Lambda}$ hyperons
at $\sqrt{s_{NN}}$ = 62.4 GeV and
200 GeV performed with the STAR detector at RHIC \cite{Abelev:2007zk} 
showed result consistent with zero within the precision of the measurements.
Later measurements by the STAR Collaboration gave nonzero values for the 
global polarization in the energy range $\sqrt{s_{NN}}$ = 7.7--200 GeV
\cite{STAR:2017ckg,Adam:2018ivw}, see Fig. \ref{fig1.2}.  
These results required more quantitative approaches to the calculation 
of the global polarization.

The most pupular now thermodynamic approach to such calculations was proposed in
Refs. \cite{Becattini:2007nd,Becattini:2013fla} and then further elaborated 
\cite{Becattini:2016gvu,Fang:2016vpj}. More precisely, the derivation of 
Refs. \cite{Becattini:2007nd,Becattini:2013fla} was performed in terms of the 
hadronic degrees of freedom and for a (1/2)-spin particle. 
The thermodynamic approach does not require the precise form of the spin-orbital coupling. 
It requires only the fact that such a coupling exists and its scale is 
the strong-interaction one. The latter is required for fast local equilibration of 
spin degrees of freedom similar to that for momentum degrees of freedom. 
Nevertheless, certain features of the spin-orbital coupling are silently assumed in 
the thermodynamic approach. In particular, this coupling is assumed to be the same 
for baryons and anti-baryons.  Had it be a coupling because of induced magnetic 
field (like in the Barnett effect), 
polarizations of baryons and anti-baryons would be of opposite sign. 
All together, this makes possible a definite quantitative estimate of polarization through
a suitable extension of the freeze-out formula.

The measured polarization is 
generally reproduced within the thermodynamics approach 
\cite{Becattini:2013fla,Becattini:2016gvu,Fang:2016vpj}
implemented in hydrodynamic UrQMD+vHLLE \cite{Karpenko:2016jyx} and 
PICR \cite{Xie:2017upb} models, as well as  
kinetic AMPT \cite{Li:2017slc,Sun:2017xhx,Wei:2018zfb,Shi:2017wpk} 
and PHSD \cite{Kolomeitsev:2018svb} models.
This reproduction indicates that our current understanding of the 
heavy-ion dynamics and, in particular, the vortical motion is compatible with 
observed polarization. However, a potential problem persists. 
The thermodynamic approach is unable to explain a big difference between 
polarization of $\Lambda$ and $\bar{\Lambda}$ at 7.7 GeV. 
A possible source of this big difference could be magnetic field induced 
in the course of the collision because it results in opposite 
polarization of baryons and anti-baryons. This possibility was studied 
in Ref. \cite{Han:2017hdi}. 
It was found found that there is no way to obtain the large
splitting of the spin polarization between $\Lambda$ and $\bar{\Lambda}$ 
at $\sqrt{s_{NN}}$ = 7.7 GeV with partonic dynamics.
It is not quite clear whether 
this contradiction between the predictions of the thermodynamics approach and 
the STAR data at energy of 7.7 GeV really takes place
because of large error bars 
of the measured $\bar{\Lambda}$ polarization. More accurate measurements are needed to 
clarify this problem.

However, there are approaches which 
naturally explain this difference. One of them is that directly based on the axial vortical effect 
\cite{Rogachevsky:2010ys,Gao:2012ix,Sorin:2016smp}. 
The axial vortical effect is associated with axial-vector current 
induced by vorticity. This current implies that the right (left)-handed
fermions move parallel (opposite) to the direction of vorticity. As the momentum of a right (left)-handed massless fermion is parallel (opposite) to its spin, all spins become
parallel to the direction of vorticity, i.e. aligned.  
Application of this approach with parameters deduced from 
the quark-gluon string transport model \cite{Baznat:2017jfj} well reproduces 
both the $\bar{\Lambda}$ and $\Lambda$ polarizations and splitting between them. Another recently 
suggested approach \cite{Csernai:2018yok} based on a Walecka-like model can also 
explain the difference in the $\bar{\Lambda}$-$\Lambda$ polarizations. However, 
ability of this Walecka-like approach to describe absolute values of these 
polarizations still remains to be seen.

The problem of relation between the vorticity and
spin-polarization tensor in relativistic systems is far from being 
completely solved. 
The thermodynamic derivation the spin polarization in Refs. 
\cite{Becattini:2007nd,Becattini:2013fla,Becattini:2016gvu} is not completely
consistent. The author themselves call it an educated ansatz \cite{Becattini:2018lge}. 
A critical review of this issue is presented in Ref. \cite{Florkowski:2018ahw}.

In the present paper we review studies of vortical motion 
and the resulting global polarization of $\Lambda$ and $\bar{\Lambda}$ hyperons
in Au+Au collisions performed within the 3FD model 
\cite{Ivanov:2017dff,Ivanov:2018eej,Ivanov:2019ern}. 
The 3FD model describes of the major part of bulk
observables: the baryon stopping \cite{Ivanov:2013wha,Ivanov:2012bh}, 
yields of different hadrons, their rapidity and transverse momentum
distributions \cite{Ivanov:2013yqa,Ivanov:2013yla}, and also  
the elliptic \cite{Ivanov:2014zqa} 
and directed \cite{Konchakovski:2014gda} flow. 
It also reproduces \cite{Ivanov:2018vpw} recent STAR data on bulk observables 
\cite{Adamczyk:2017iwn}. 
We also address the question why the observed global polarization of 
hyperons in the midrapidity region, i.e. at pseudorapidity $|\eta|<$ 1 (see Fig. \ref{fig1.2}), 
drops with the collision energy rise  
while the angular momentum 
accumulated in the system substantially increases (see Fig. \ref{fig1}) 
at the same time?

\section{Vorticity in the 3FD model}
\label{Results}

The  3FD model takes into account 
a finite stopping power resulting in counterstreaming 
of leading baryon-rich matter at early stage of nuclear collisions 
\cite{3FD}. This 
nonequilibrium stage 
is modeled by means of two counterstreaming baryon-rich fluids 
initially associated with constituent nucleons of the projectile
(p) and target (t) nuclei. 
Later on these  fluids may consist
of any type of hadrons and/or partons (quarks and gluons),
rather than only nucleons.
Newly produced particles, dominantly
populating the midrapidity region, are associated with a fireball
(f) fluid.
These fluids are governed by conventional hydrodynamic equations 
coupled by friction terms in the right-hand sides of the Euler equations. 
The friction results in energy--momentum loss of the 
baryon-rich fluids. A part of this
loss is transformed into thermal excitation of these fluids, while another part 
leads to formation of the fireball fluid.
Thus, the 3FD approximation is a minimal way to implement the early-stage nonequilibrium 
of the produced strongly-interacting matter at high collision energies.

Three different 
equations of state (EoS's) were used in simulations of Refs. 
\cite{Ivanov:2013wha,Ivanov:2012bh,Ivanov:2013yqa,Ivanov:2013yla,Ivanov:2014zqa,Konchakovski:2014gda,Ivanov:2018vpw}: a purely hadronic EoS \cite{gasEOS}  
and two versions of the EoS with the   deconfinement
 transition \cite{Toneev06}, i.e. a first-order phase transition  
and a crossover one. 
In the present review  
only the first-order-phase-transition (1PT) and crossover EoS's 
are discussed as the most relevant to various observables.

A so-called thermal vorticity is defined as 
   \begin{eqnarray}
   \label{therm.vort.}
   \varpi_{\mu\nu} = \frac{1}{2}
   (\partial_{\nu} \hat{\beta}_{\mu} - \partial_{\mu} \hat{\beta}_{\nu}), 
   \end{eqnarray}
where 
$\hat{\beta}_{\mu}=\hbar\beta_{\mu}$,  $\beta_{\mu}=u_{\nu}/T$, 
$u_{\mu}$ is collective local four-velocity of the matter,  and
$T$ is local temperature.  
In the thermodynamical approach \cite{Becattini:2013fla,Becattini:2016gvu,Fang:2016vpj}
in the leading order in the thermal vorticity 
it is directly related to 
the mean spin vector of spin 1/2 particles with four-momentum $p$, 
produced around point $x$ on freeze-out hypersurface 
   \begin{eqnarray}
\label{xp-pol}
 S^\mu(x,p)
 =\frac{1}{8m}     [1-n_F(x,p)] \: p_\sigma \epsilon^{\mu\nu\rho\sigma} 
  \varpi_{\rho\nu}(x) 
   \end{eqnarray}
where $n_F(x,p)$ is the Fermi-Dirac distribution function and $m$ is mass of the 
considered particle. 
To calculate the
relativistic mean spin vector of a given particle species
with given momentum, the above expression should be integrated
over the freeze-out hypersurface. Therefore, we proceed to discussion in terms of 
the thermal vorticity.

Unlike the conventional hydrodynamics,  
the system is characterized by three hydrodynamical
velocities, $u^\mu_a$ ($a$ = p, t and f), in the 3FD model. 
The counterstreaming of
the p and t fluids takes place only at the initial stage of
the nuclear collision  
that lasts from $\sim$ 5 fm/c at $\sqrt{s_{NN}}=$ 5 GeV \cite{Ivanov:2017dff}
to $\sim$ 1 fm/c at collision energy of 39 GeV \cite{Ivanov:2017xee}. 
At later stages the baryon-rich (p and t) fluids have already 
either partially passed though each other or 
partially stopped and unified in the central region. 
Therefore, after the initial thermalization stage the system is characterized 
 by two hydrodynamical velocities, $u^\mu_{\rm B}$ and $u^\mu_{\rm f}$, 
and two temperatures, $T_{\rm B}$ and $T_{\rm f}$, corresponding to the 
unified baryon-rich (B) and fireball (f) fluids.

\begin{figure}[!htb]
\includegraphics[width=5.cm]{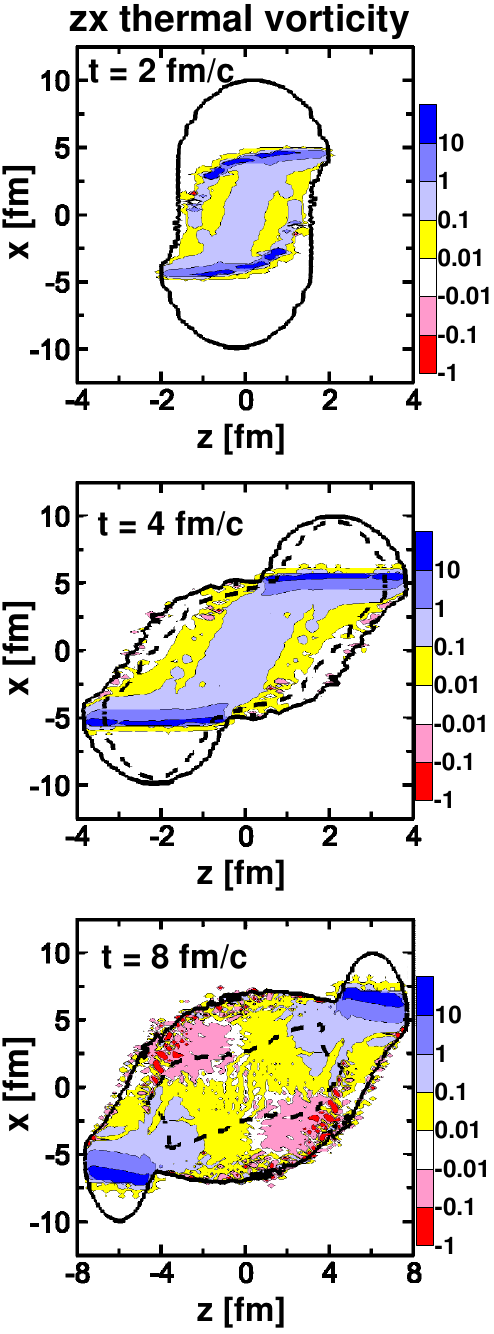}
 \caption{
The proper-energy-density weighted 
thermal $zx$ vorticity 
of the baryon-rich subsystem,  
in the reaction plane at various time instants 
in the semi-central ($b=$ 6 fm) Au+Au collision at $\sqrt{s_{NN}}=$ 7.7 GeV. 
Calculations are done with the crossover EoS. $z$ axis is the 
beam direction. Note different scale along the $z$ axis at different time instants. 
The outer bold solid contour displays the border of the baryon-rich matter. Inside this contour 
$n_B/n_0 > 0.1$.  
The inner bold dashed contour indicates the freeze-out border. Inside this contour 
the matter still hydrodynamically evolves, while outside -- it is frozen out.
At $t=$ 2 and 4 fm/c there is no frozen-out matter.
}
\label{fig2}
\end{figure}

As a result the system is characterized by 
two sets of the  vorticity related to these baryon-rich   
and baryon-free fluids, 
$\varpi_{\mu\nu}^{\rm B}$ and $\varpi_{\mu\nu}^{\rm f}$, respectively. 
In order to define define a single 
quantity responsible for the particle polarization
we make sum of these vorticities  
with the weights of their energy densities  
%
   \begin{eqnarray}
   \label{en.av.rel.B-vort}
   \widetilde{\varpi}_{\mu\nu} ({\bf x},t) 
   =  
   \frac{\varpi_{\mu\nu}^{\rm B}({\bf x},t) \varepsilon_{\rm B} ({\bf x},t)
   + 
   \varpi_{\mu\nu}^{\rm f}({\bf x},t)  \varepsilon_{\rm f} ({\bf x},t)}{ 
 \varepsilon ({\bf x},t)}   
   \end{eqnarray}
where $\varepsilon_{\rm B}$ and $\varepsilon_{\rm f}$ are the proper (i.e. in the local rest frame) 
energy densities of  
the baryon-rich and baryon-free fluids, respectively, and  $\varepsilon$ is 
proper energy density of all three fluids in their combined local rest frame.  
In view of almost perfect unification of the baryon-rich fluids 
and small local baryon-fireball relative velocities \cite{Ivanov:2017xee} 
at the later stages of the collision,
a very good approximation for $\varepsilon$ is just 
\begin{eqnarray}
\label{eps-tot-appr}
\varepsilon \simeq  \varepsilon_{\scr B} + \varepsilon_{\scr f}. 
\end{eqnarray}
The proper-energy-density weighted  vorticity allows us 
to suppress contributions of regions of low-density matter.  
It is appropriate because production of (anti)hyperons under consideration 
dominantly takes place in highly excited regions of the system.

In Fig.  \ref{fig2}, the proper-energy-density weighted 
thermal $zx$ vorticity of the baryon-rich subsystem 
in the reaction plain ($xz$) is presented
at various time instants 
in semi-central ($b=$ 6 fm) Au+Au collisions at $\sqrt{s_{NN}}=$ 7.7 GeV. 
As seen, the thermal vorticity
primarily starts at the border between the participant and spectator matter. 
Later on it partially spreads to the participant and spectator bulk though 
remain concentrated near the border.  
In the conventional hydrodynamics this 
extension into the bulk of the system is an effect of the shear viscosity. In the 3FD dynamics it 
is driven by the 3FD dissipation which imitates the effect of the shear viscosity \cite{Ivanov:2016vkw}.
The spread into the bulk, i.e. into the midrapidity region, 
is stronger at lower collision energies  \cite{Ivanov:2017dff}
because of the higher effective shear viscosity than that at higher energies \cite{Ivanov:2016vkw}. 
This explains the drop of the vorticity value and consequently 
the observed hyperon polarization at the midrapidity with the collision energy rise.

At later times the maximum values in the vortical fields get more and more shifted 
to the fragmentation regions because of the 1D expansion of the system. 
At the same time, the vorticity in the participant bulk gradually dissolves.   
It is peculiarly that four strong oppositely directed vortices are formed at the periphery 
of the fragmentation regions, see Fig. \ref{fig2}.   
The vortex at the border with the spectator matter is an order of magnitude stronger 
than its counterpart. This is the structure as it is seen in the reaction plane.

\begin{figure}[!htb]
\includegraphics[width=8.cm]{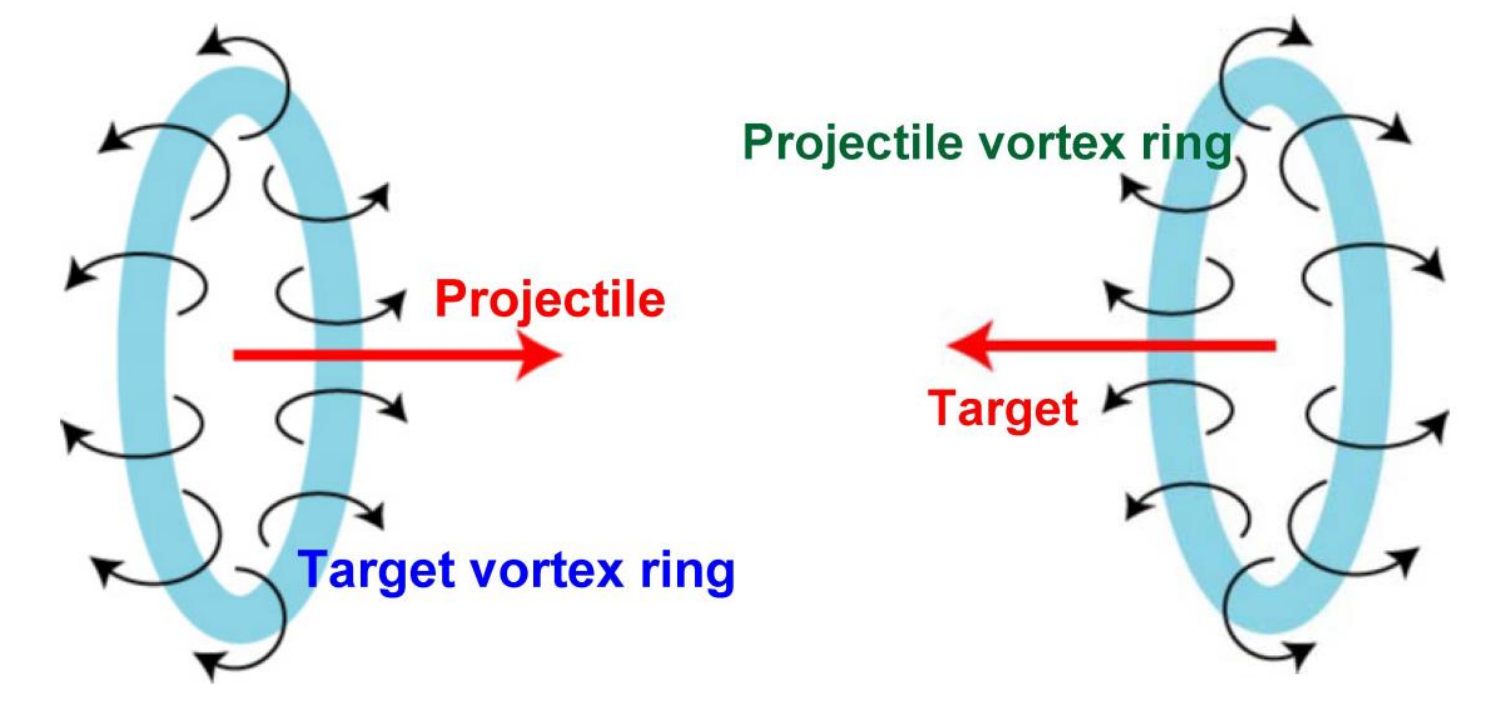}
 \caption{%
Schematic picture of the vortex rings in the fragmentation regions. 
Curled arrows indicate direction of the circulation of the target matter. 
}
\label{fig3}
\end{figure}

In fact, in three dimensions these are two vortex rings: one in the target fragmentation region and another in 
the projectile one. The matter rotation is opposite in these two rings. They are 
formed because the matter in the vicinity of the beam axis ($z$) is stronger decelerated 
 because of thicker matter in the center than that at the periphery. 
Indeed, these rings are formed at the transverse periphery 
of the stopped matter in the central region, see the central bumps in $n_B$ and $\varepsilon$ 
at $t=$ 1 fm/c in Fig. \ref{fig2}. 
Thus, the peripheral matter acquires a rotational motion. 
A schematic picture of these vortex rings in the fragmentation regions
is presented in Fig. \ref{fig3}.  

A similar effect was noticed in the analysis of the vorticity field 
\cite{Baznat:2015eca,Baznat:2013zx} at lower NICA energies. 
The authors of  Refs. \cite{Baznat:2015eca,Baznat:2013zx} called this 
specific toroidal structure as a femto-vortex sheet. 
This femto-vortex sheet is not a ring because the vorticity disappears 
in the $xy$ plane, i.e. in the plane orthogonal to the reaction $xz$ plane. 
At higher collision energies this femto-vortex sheet splits into two real 
rings. 

\begin{figure}[!htb]
\includegraphics[width=4.9cm]{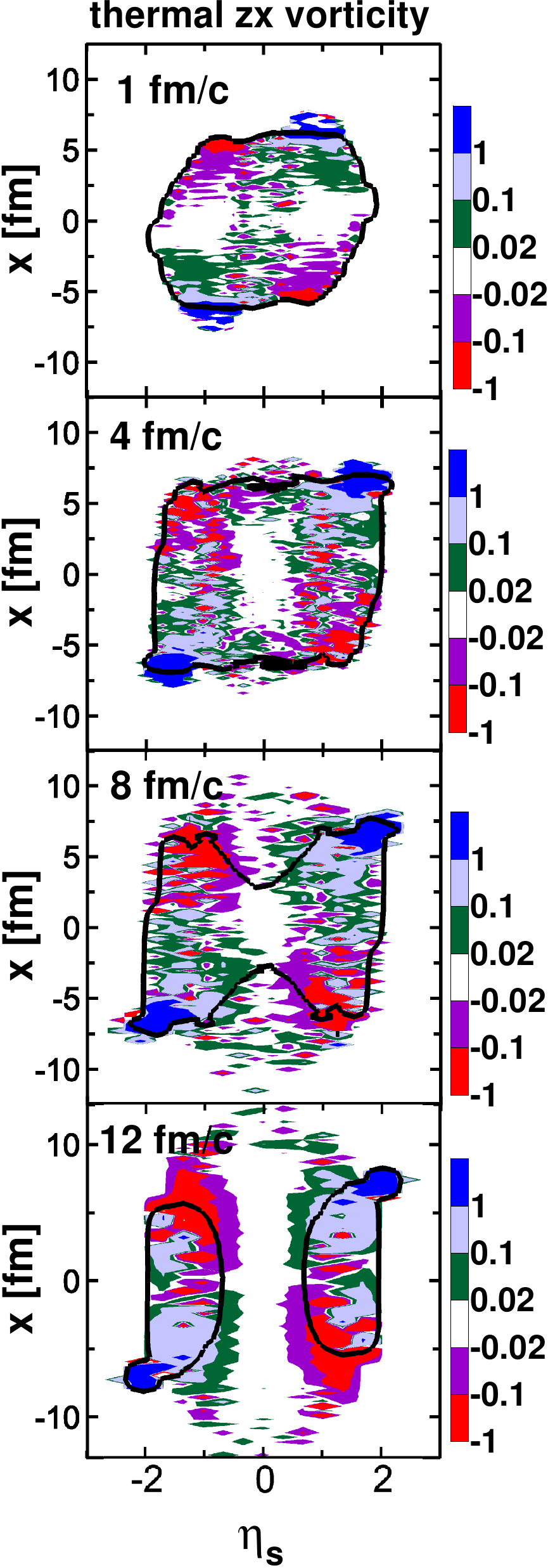}
 \caption{%
The proper-energy-density weighted 
 thermal $zx$ vorticity 
in the central ($b=$ 2 fm) Au+Au collision at $\sqrt{s_{NN}}=$ 39 GeV. 
$\eta_s$  is the space-time rapidity along the beam direction, see Eq. (\ref{eta_s}).
Calculations are done with the 1PT EoS. 
The bold solid contour displays the border of the frozen out matter.  Inside this contour 
the matter still hydrodynamically evolves, while outside -- it is frozen out.
}
\label{fig4}
\end{figure}

These rings are also formed in central collisions, as seen from Fig. \ref{fig4}. 
As seen, the vortex rings are formed already at $t=$ 1 fm/c. 
In fact, the  schematic picture of the completely symmetric vortex ring, 
see Fig. \ref{fig3}, corresponds to the exactly central collision at $b=$ 0.

Figure \ref{fig4} presents the time evolution of the proper-energy-density weighted 
 thermal $zx$ vorticity 
in the reaction plain ($x\eta_s$) of
central Au+Au collision at $\sqrt{s_{NN}}=$ 39 GeV, where 
   \begin{eqnarray}
   \label{eta_s}
   \eta_s = \frac{1}{2} \ln\left(\frac{t+z}{t-z}\right)
   \end{eqnarray}
is the longitudinal space-time rapidity and $z$ is the coordinate along the beam direction. 
The advantage of this longitudinal space-time rapidity is that it zooms Lorentz-contracted regions.

\section{Polarization}
\label{polarization}

The above observation of pushing out the vortical field to the fragmentation regions 
has consequence for the  polarization of produced particles. 
The polarized particles dominantly originate from peripheral regions with high vorticity, 
see right panel in Fig. \ref{fig2} and Fig. \ref{fig4}. 
Therefore, the relative polarization of $\Lambda$ hyperons
should be higher in the fragmentation regions, 
i.e. the kinematical region of the participant-spectator border,   
than that in the midrapidity region.

In terms of the mean spin vector (\ref{xp-pol}), 
the polarization vector of $S$-spin particle is defined as 
   \begin{eqnarray}
   \label{P_S}
  P^\mu_{S} = S^\mu / S.  
   \end{eqnarray}
In the experiment, the  polarization of the $\Lambda$ hyperon is measured in
its rest frame, therefore the $\Lambda$ polarization is 

   \begin{eqnarray}
   \label{P_L-rest}
  P^\mu_{\Lambda} = 2 S^{*\mu}_{\Lambda}  
   \end{eqnarray}
where $S^{*\mu}_{\Lambda}$ is mean spin vector of the $\Lambda$ hyperon in its rest frame. 
Substituting expression for ${\bf S}$ from Eq. (\ref{xp-pol}) and averaging it 
over the ${\bf p}_{\Lambda}$ direction (i.e. over ${\bf n}_p$) 
we arrive at the following 
polarization along the $y$ axis 
\cite{Kolomeitsev:2018svb}
(see also \cite{Becattini:2013fla,Becattini:2016gvu,Fang:2016vpj}) 
   \begin{eqnarray}
   \label{P_Lambda}
\langle  P_{\Lambda}\rangle_{{\bf n}_p} =  
 \frac{1}{2m_{\Lambda}}
 \left(E_{\Lambda} - \frac{1}{3} \frac{{\bf p}_{\Lambda}^2}{E_{\Lambda}+m_{\Lambda}} \right)
 \varpi_{zx},  
   \end{eqnarray}
where $m_{\Lambda}$  is the $\Lambda$ mass, 
$E_{\Lambda}$ and ${\bf p}_{\Lambda}$ are the energy and momentum of the emitted $\Lambda$ hyperon, respectively. 
Here we put $(1-n_\Lambda) \simeq 1$ because the $\Lambda$ production takes place only 
in high-temperature regions, where Boltzmann statistics dominates.

\begin{figure}[bht]
\includegraphics[width=7.cm]{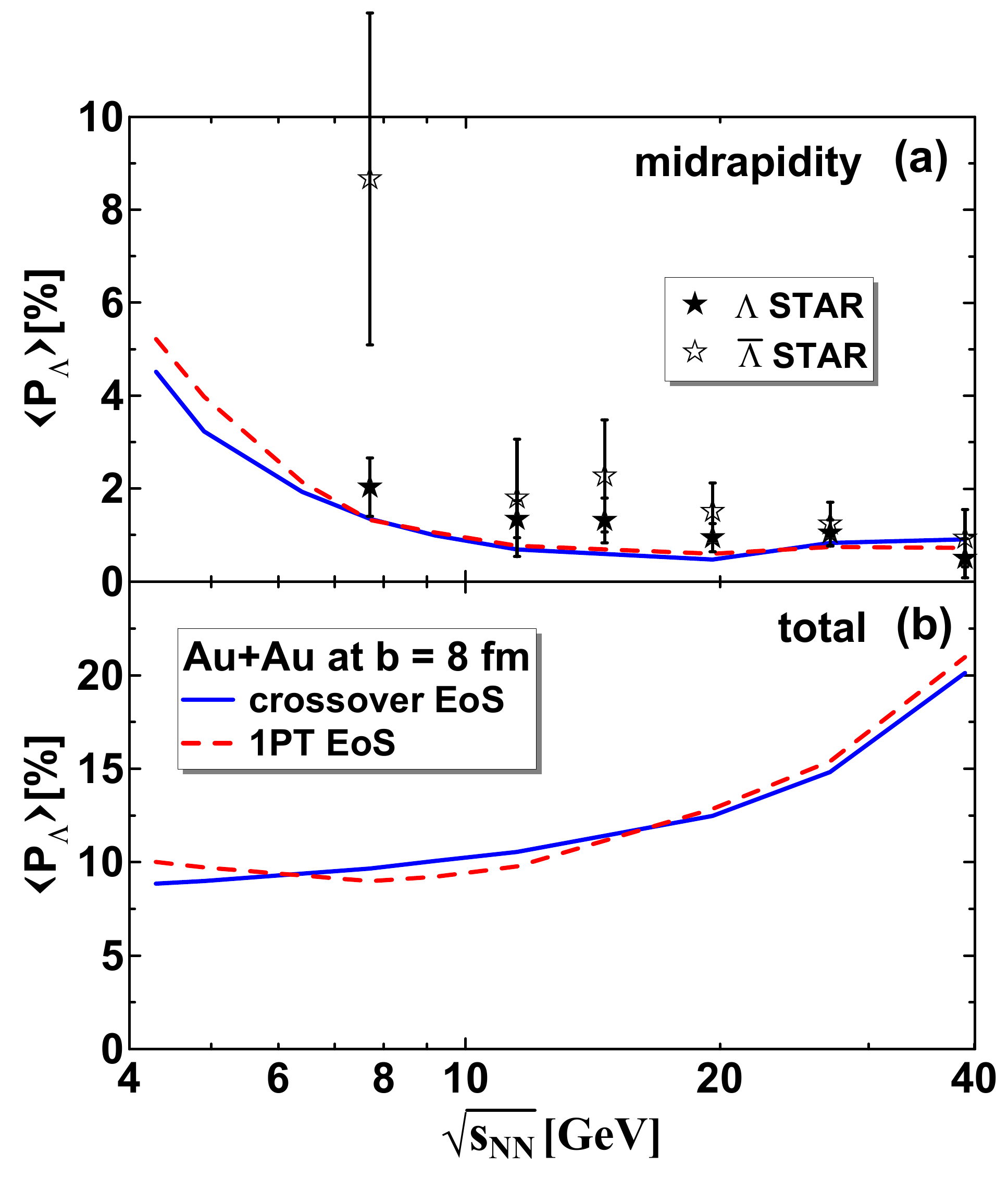}
 \caption{
 Global (a), i.e. in the central region, and total (b), i.e. averaged over the whole participant region,
polarization of $\Lambda$ hyperons in Au+Au collisions at $b=$ 8 fm as a function of collision energy $\sqrt{s_{NN}}$. 
STAR data on global $\Lambda$ and also $\bar{\Lambda}$ polarization in the midrapidity 
region (pseudorapidity cut $|\eta|<$ 1) \cite{STAR:2017ckg} are also displayed. 
}
\label{fig6}
\end{figure}

We apply further approximations after which  the present evaluation of the global
polarization becomes more an estimation rather than a calculation. 
We associate the global midrapidity polarization with the polarization of $\Lambda$ hyperons 
emitted from the central region (i.e. central slab) 
of colliding system, see details in Ref. \cite{Ivanov:2019ern}. 
We decouple averaging of $\varpi_{zx}$ and the term in parentheses in Eq. (\ref{P_Lambda}). 
Neglecting the longitudinal motion of the $\Lambda$ hyperon  
in the central slab, 
we approximate the average $\Lambda$ energy by the mean midrapidity transverse mass: 
$\langle E_{\Lambda}\rangle = \langle m_T^\Lambda \rangle_{\scr{midrap.}}$,
which was calculated earlier in Ref. \cite{Ivanov:2013yla}. 
Thus we arrive at the  
 estimate of the global 
midrapidity $\Lambda$-polarization in the $y$ direction 
   \begin{eqnarray}
   \label{mid.r.-P_Lambda}
  \langle P_{\Lambda} \rangle_{\scr{midrap.}} \simeq 
 \frac{\langle \varpi_{zx} \rangle_{\scr{cent. slab}} }{2}
 \left(1 +  \frac{2}{3} 
 \frac{\langle m_T^\Lambda \rangle_{\scr{midrap.}} - m_\Lambda}{m_{\Lambda}} \right) 
\cr
   \end{eqnarray}
Results of this estimate are presented in panel (a) of Fig. \ref{fig6}. 
The corresponding 3FD simulations of Au+Au collisions were performed at fixed
impact parameters $b=$ 8 fm. This value of $b$ was chosen in order to roughly 
comply with the centrality selection 20-50\% in the STAR experiment \cite{STAR:2017ckg}. 
The correspondence between experimental centrality and the mean impact parameter
was taken from Glauber simulations of Ref. \cite{Abelev:2008ab}.

%
%

As seen from Fig. \ref{fig6}, such a rough estimate of the global midrapidity polarization
quite satisfactorily reproduces the experimental data, especially the collision-energy 
dependence of the polarization. This energy dependence is related to the above discussed 
decrease of the thermal vorticity in the central region with the collision 
energy rise. The performed estimate predicts that the global midrapidity polarization 
further increases at NICA/FAIR energies, reaching values of 5\% at $\sqrt{s_{NN}}=$ 3.8 GeV. 
This prediction approximately agrees with that made in Ref. \cite{Baznat:2017jfj} 
based on the axial vortical effect \cite{Rogachevsky:2010ys,Gao:2012ix,Sorin:2016smp}.

The global midrapidity polarization of $\bar{\Lambda}$ hyperons differs only by  
replacement of $\langle m_T^{\Lambda} \rangle_{\scr{midrap.}}$ by 
$\langle m_T^{\bar{\Lambda}} \rangle_{\scr{midrap.}}$  
in Eq. (\ref{mid.r.-P_Lambda}) from that for $\Lambda$'s
and quantitatively does not exceed 5\% of that for $\Lambda$ hyperons. Therefore, 
we do not display it in Fig. \ref{fig6}. Uncertainties of this estimate were 
discussed in Ref. \cite{Ivanov:2019ern}. These were found to be of the order of 
20\%, of course excluding the main uncertainty associated with decoupling of 
averaging of $\varpi_{zx}$ and the term in parentheses in Eq. (\ref{P_Lambda}).

In the case of total $\Lambda$ polarization, the averaging  
runs over the whole participant region.
In such averaging the above applied 
 decoupling of averaging of $\varpi_{zx}$ and the term in parentheses in Eq. (\ref{P_Lambda}) 
is even less justified than in the central region. Therefore, we do 
even more rough estimate of the
mean total polarization of emitted $\Lambda$ hyperons 
%
   \begin{eqnarray}
   \label{mean-P_Lambda}
  \langle P_{\Lambda} \rangle_{\rm total} \approx  
 \frac{\langle \varpi_{zx} \rangle}{2}
     \end{eqnarray}
by neglecting the term in parentheses in Eq. (\ref{P_Lambda}). Note that this term 
is a correction, though not a negligible one. Sometimes it results in 
30\% correction for the central-slab polarization.

Results of this estimate of the total $\Lambda$ polarization
are presented in panel (b) of Fig. \ref{fig6}. 
The total $\Lambda$ polarization increases with collision energy rise. 
This is in contrast to the energy dependence of the midrapidity polarization. 
This increase is quite moderate as compared with the rapid rise of the 
angular momentum accumulated in the participant region, see Fig. \ref{fig1}. 
The increase of the total polarization with simultaneous decrease of the 
midrapidity one suggests that the $\Lambda$ polarization in the fragmentation regions 
reaches high values at high collision energies. 
At lower collision energies values of the total and midrapidity polarization are close 
to each other, which reflects a more homogeneous distribution of the vortical field over 
the bulk of the produced matter.

\section{Summary}
\label{Summary}

We reviewed studies of the vortical motion 
and the resulting global polarization of $\Lambda$ and $\bar{\Lambda}$ hyperons
in heavy-ion collisions, in particular, within the 3FD model 
\cite{Ivanov:2017dff,Ivanov:2018eej,Ivanov:2019ern}. 
The measurement of the polarization give us possibility to deduce information 
on the vortical motion of the matter produced in heavy-ion collisions.  
The interrelation between the vortical motion and the polarization is not quite 
clear at present. 
Calculations of the global polarization of $\Lambda$ and $\bar{\Lambda}$ hyperons
based on the thermodynamics approach 
\cite{Becattini:2013fla,Becattini:2016gvu,Fang:2016vpj}
implemented within various models 
\cite{Karpenko:2016jyx,Xie:2017upb,Li:2017slc,Sun:2017xhx,Wei:2018zfb,Shi:2017wpk,Kolomeitsev:2018svb,Ivanov:2019ern}, 
including the 3FD one, are consistent with data by the STAR Collaboration  in the energy range $\sqrt{s_{NN}}$ = 7.7--200 GeV \cite{STAR:2017ckg,Adam:2018ivw}. 
However, the thermodynamic approach is unable to explain a big difference between 
polarization of $\Lambda$ and $\bar{\Lambda}$ at 7.7 GeV. 
It should be mentioned that 
there are altenative approaches 
\cite{Rogachevsky:2010ys,Gao:2012ix,Sorin:2016smp,Baznat:2017jfj,Csernai:2018yok},  
which naturally explain this difference. 
In view of large error bars 
of the measured $\bar{\Lambda}$ polarization at this energy, more accurate measurements are needed to 
clarify this problem.

Based on the analysis within the 3FD model \cite{Ivanov:2019ern}
predictions are made for  
the global midrapidity polarization in the FAIR-NICA 
\cite{Friman:2011zz,Kekelidze:2017ghu} energy range. 
The performed estimate predicts that the global midrapidity polarization 
further increases at NICA/FAIR energies, reaching values of 5\% at $\sqrt{s_{NN}}=$ 3.8 GeV. 
This prediction approximately agrees with that made in Ref. \cite{Baznat:2017jfj} 
based on the axial vortical effect \cite{Rogachevsky:2010ys,Gao:2012ix,Sorin:2016smp}.

It is found that the energy dependence the observed global polarization of 
hyperons in the midrapidity region 
is a consequence of the decrease 
of the thermal vorticity in the central region with the collision 
energy rise, which in its turn results from pushing out the vorticity field into  
the fragmentation regions \cite{Ivanov:2018eej,Jiang:2016woz}.

At high collision energies $\sqrt{s_{NN}}\agt$ 8 GeV, this pushing-out 
results in a peculiar vortical structure consisting of two vortex rings: 
one ring in the target fragmentation region and another one in 
the projectile fragmentation region 
with the matter rotation being opposite in these two rings
\cite{Ivanov:2018eej}, see Fig. \ref{fig3}. 
These vortex rings produce very strong $\Lambda$ polarization in 
the fragmentation regions at noncentral collisions.

\vspace*{2mm} {\bf Acknowledgments} \vspace*{2mm}


The problems considered in this contribution have been repeatedly discussed with E. E. Saperstein. 
We are very grateful to him for these discussions.
Fruitful discussions with E. E. Kolomeitsev and O. V. Teryaev are gratefully acknowledged.
This work was carried out using computing resources of the federal collective usage center ``Complex for simulation and data processing for mega-science facilities'' at NRC "Kurchatov Institute", http://ckp.nrcki.ru/.
Y.B.I. was supported by the Russian Science
Foundation, Grant No. 17-12-01427, and the Russian Foundation for
Basic Research, Grants No. 18-02-40084 and No. 18-02-40085. 
A.A.S. was partially supported by  the Ministry of Education and Science of the Russian Federation within  
the Academic Excellence Project of 
the NRNU MEPhI under contract 
No. 02.A03.21.0005. 

\end{document}